# Spin-Dependent Graph Neural Network Potential for Magnetic Materials

Hongyu Yu[a,b], Yang Zhong[a,b], Liangliang Hong[a,b], Changsong Xu[a,b], Wei Ren,[c] Xingao Gong,[a,b] Hongjun Xiang[a,b,*]

[a]*Key Laboratory of Computational Physical Sciences (Ministry of Education), Institute of Computational Physical Sciences, State Key Laboratory of Surface Physics, and Department of Physics, Fudan University, Shanghai 200433, China*
[b]*Shanghai Qi Zhi Institute, Shanghai 200030, China*
[c]*Department of Physics, International Center of Quantum and Molecular Structures and Materials Genome Institute, Shanghai University 200444, Shanghai, China*

**Abstract**:

The development of machine learning interatomic potentials has immensely contributed to the accuracy of simulations of molecules and crystals. However, creating interatomic potentials for magnetic systems that account for both magnetic moments and structural degrees of freedom remains a challenge. This work introduces SpinGNN, a spin-dependent interatomic potential approach that employs the graph neural network (GNN) to describe magnetic systems. SpinGNN consists of two types of edge GNNs: Heisenberg edge GNN (HEGNN) and spin-distance edge GNN (SEGNN). HEGNN is tailored to capture Heisenberg-type spin-lattice interactions, while SEGNN accurately models multi-body and high-order spin-lattice coupling. The effectiveness of SpinGNN is demonstrated by its exceptional precision in fitting a high-order spin Hamiltonian and two complex spin-lattice Hamiltonians with great precision. Furthermore, it successfully models the subtle spin-lattice coupling in $BiFeO_3$ and performs large-scale spin-lattice dynamics simulations, predicting its antiferromagnetic ground state, magnetic phase transition, and domain wall energy landscape with high accuracy. Our study broadens the scope of graph neural network potentials to magnetic systems, serving as a foundation for carrying out large-scale spin-lattice dynamic simulations of such systems.

# Introduction

Machine learning (ML) interatomic potentials have transformed the field of computational condensed matter physics by enabling highly accurate simulations of materials [1,2]. These potentials are trained on a dataset of atomic configurations and their corresponding total energies or forces using a machine-learning model. This allows the model to learn the potential energy surface (PES) of the system and simulate its behavior under different conditions such as temperature, pressure, or deformation. Compared to traditional methods like density functional theory, ML interatomic potentials are much faster, making them a powerful tool for investigating large and complex materials systems over long timescales[3–7]. Kernel-based ML interatomic potentials employ material descriptors as inputs and learn to map them to the corresponding energy of the material. These methods include kernel ridge regression (KRR) and Gaussian process regression (GPR). The descriptors, which are derived from the atomic positions and atomic numbers, remain invariant under uniform translations, rotations, and permutations of identical atoms in the material. The construction of the descriptors significantly affects the performance of the MLIP model. Various descriptors have been developed to encode material information such as Behler-Parrinello neural networks[8], GAP[9], SNAP[10], Moment Tensor Potentials[11], DeepMD[12], ACE[13], SOAP[14], and ACSFs[15].

**Graph neural network and Message-passing neural network**

Graph neural networks (GNNs) with message-passing deep learning architecture have shown remarkable performance in condensed matter physics. Message-passing neural network (MPNN) interatomic potentials induce an atomistic graph by connecting each atom (node) to neighboring atoms within a finite cutoff sphere. This structure representation is learned directly from the input structure, which makes it more natural than manual descriptors construction methods. The graph representations of GNN naturally include symmetry invariants such as translation, rotation, and permutation. Recently, molecular and crystal graph neural networks have been well-developed as end-to-end learned natural descriptors for crystals[16–23]. The Line Graph Neural Network (LGNN), a recently proposed extension of GNN, explicitly includes bond angles and updates edge features in the graph, exemplified by ALIGNN[24], and Dimenet++[19]. Furthermore, equivariant neural networks have been developed to process non-invariant geometric inputs, such as displacement vectors, while maintaining symmetry. By using only E(3)-equivariant operations, these models achieve internal features that are equivariant with respect to the 3D Euclidean group. Equivariant architectures provide an approach for creating interatomic potential models such as NequIP[22] and MACE[25]. In particular, the NequIP model, along with other equivariant implementations[26–30], has demonstrated remarkable performance in accurately describing the structural and kinetic properties of complex materials, achieving unprecedented error rates across a broad range of systems and remarkable sample efficiency. The MPNN has shown outstanding performance in training potential, interpreting materials, predicting properties, and inverse design[16,17,31,32]. While MPNNs can learn many-body correlations and access non-local information beyond the local cutoff, their parallelism presents significant limitations for large-scale system simulations. The iterative propagation of information in MPNNs results in large receptive fields with many effective neighbors per atom, hindering parallel computation and restricting the accessible length scales for atom-centered message-passing machine learning interatomic potentials (MLIPs). A potential solution to these challenges is Allegro[23], a recently proposed interatomic potential architecture that uses strictly local equivariant deep neural networks for scalable and accurate modeling. As a result, molecular dynamics simulations can now predict the structural and kinetic properties of complex systems containing millions of atoms with near-first-principles fidelity.

**Magnetic interatomic potential**

In magnetic materials, the potential energy surface (PES) depends not only on the positions of the atoms but also on the spin configurations of the magnetic atoms. At finite temperatures, atomic magnetic moments fluctuate, and a single magnetic state is insufficient to describe the system. Even at ambient temperatures, atomic spin flips occur frequently due to small energy differences between different spin configurations, typically in the range of a few meV/atom. Therefore, a thorough understanding of magnetic materials necessitates incorporating the effects of temperature on atomic spin configurations. The effective spin-lattice Hamiltonian is traditionally used to understand the spin-lattice coupling in multiferroic materials, such as $BiFeO_3$, based on the system's symmetry[33–36]. However, it is challenging to determine the terms and coefficients required to build the potential, and the effective Hamiltonian method is limited in its ability to describe systems with significant perturbations from the ideal structure.

MLIPs have advanced significantly in their ability to investigate complex systems and physical phenomena, but their failure to consider the diverse range of spin arrangements and magnetic interactions remains a limitation. Many descriptors only work with potentials that do not include atomic magnetic moments, which makes most MLIP models unable to describe the intricate spin-lattice coupling present in magnetic materials. Representing magnetic PES is even more challenging than non-magnetic PES due to the dimensional crisis. For illustration, for a system with N particles, a non-magnetic potential function has 3×N freedom degrees as input of $E = f(\{\vec{r_i}\})$, while for a magnetic potential function, it has 6×N freedom degrees, including atomic positions and magnetic moments as $E = f(\{\vec{r_i}\}, \{\vec{s_i}\})$, which makes it much more complicated to describe a magnetic system well. Also, one significant limitation is the cost involved in generating a dataset that is well-sampled in $\{\vec{r_i}, \vec{s_i}\}$ configuration spaces with high accuracy. The spin-polarized first-principle calculation is much more expensive than the ones without spin-polarized. Consequently, there is a significant need for a general and robust magnetic neural network potential with high accuracy and efficiency to investigate and better understand magnetic materials.

Magnetic Moment Tensor Potentials (mMTPs) [37], developed by Ivan et al., incorporate the contributions of collinear magnetic moments on Moment Tensor Potentials (MTPs) to add collinear magnetic degrees of freedom to MLIP. Although mMTPs have been demonstrated for bcc iron with calculations and simulations, it should be noted that they can only handle collinear spin and do not explicitly consider spin-inversion symmetry. Marco et al. developed the magnetic high dimensional neural network potential (mHDNNP), which modifies the ACSFs to spin-dependent ACSFs (sACSFs)[38]. However, like mMTPs, mHDNNP can only handle collinear magnetic moments, hindering the study of noncollinear magnetic states like spirals, skyrmions, and bimerons. Note that considering noncollinear magnetic states is indispensable for calculating magnetic transition temperature with Monte Carlo simulation and performing spin-lattice dynamics [39,40]. Meanwhile, Svetoslav et al. proposed Magneto-elastic ML-IAP[41] with SNAP[10] potential and a spin Hamiltonian considering Heisenberg and biquadratic terms, mainly considering the spin-spiral state. While Magneto-elastic ML-IAP can deal with a noncollinear magnetic state, the spin-lattice part is the traditional explicit term, and the complex interaction cannot be described well under this framework. Moreover, in a recent study, Behnam et al. [42] discovered that both SOAP and ACSFs, the most commonly used descriptors, fail to adequately handle four-body interactions, leading to a failure to learn the full energy landscape of materials. To date, the freedom of magnetic moments has not been included in GNN before our work, resulting in underdeveloped MLIP for noncollinear magnetic degrees of freedom.

In this work, we introduce SpinGNN, a generalized spin-dependent graph neural network framework designed to incorporate noncollinear spin into graph neural networks (GNNs). This framework extends molecular and crystal GNNs to encompass both structural and magnetic states. In our study, we find that Spin-Dimenet++ and Spin-Allegro show high accuracy and effectively capture fine spin-lattice interactions in some magnetic datasets. We further demonstrate that Spin-Allegro is not only efficient in parallelizing across devices for spin-lattice simulations that involve thousands of atoms, but it can also predict magnetic ground states accurately. Finally, we show that Spin-Allegro can predict consistent magnetic phase transitions, related transition temperatures, and energy landscape of domain walls of $BiFeO_3$, a commonly studied multiferroic material.

This article follows the subsequent structure: First, we review related work on kernel-based interatomic potential, graph neural network interatomic potential, and magnetic interatomic potential. Next, we present the core concepts of the SpinGNN framework, as well as the detailed design of Spin-Dimenet++ and Spin-Allegro based on the popular Dimenet++ and Allegro. Then, we introduce the edge-vector-based atomic virial method to enable parallel stress calculation for Allegro. Following that, we present a series of results from magnetic datasets. Finally, we demonstrate the results of spin-lattice dynamics simulation for the multiferroic $BiFeO_3$ on a large scale.

## Results

The following introduces our proposed novel method for learning magnetic potential energy surfaces, utilizing a spin-dependent graph neural network architecture that incorporates noncollinear spin with a stress parallel calculation method for Allegro.

**The SpinGNN framework**

The SpinGNN framework, illustrated in Fig. 1, enhances the standard graph neural network (GNN) potential by integrating the degrees of freedom related to atomic positions and collinear or noncollinear magnetic moments. The framework involves two separate neural network architectures, namely the Heisenberg edge graph neural network (HEGNN) and the spin-distance edge graph neural network (SEGNN). HEGNN uses updated edge features in GNN to map the variational Heisenberg coefficients between magnetic atoms. Conversely, SEGNN utilizes both the atomic distance and the dot product of the noncolinear spin between two atoms to initialize the edges in the network. By incorporating HEGNN and SEGNN in a GNN potential, we can create and train a magnetic potential model for describing the magnetic material's complicated PES. While HEGNN supports the discovery of the fundamental Heisenberg magnetic potential, SEGNN enables the modeling of complex high-order magnetic potentials. We can use both models independently or collectively to develop the magnetic potential of an ensemble. HEGNN is proficient in representing uncomplicated Heisenberg 2-order magnetic interactions, whereas SEGNN can generally illustrate both 2-order and higher-order interactions through graph convolution in GNN, rendering it competent in accommodating most spin-lattice interactions. Since the Heisenberg interactions are mostly prevalent in magnetic potentials, HEGNN is adept at constructing the magnetic potential landscape for materials with simple magnetic interactions that can be expressed through the Heisenberg model. Conversely, SEGNN should be utilized for modeling the magnetic potential landscape for materials with high-order magnetic interactions, where the Heisenberg model falls short. Hence, the choice of the SpinGNN model for creating the magnetic potential landscape is determined by the type of magnetic interactions dominating the material of interest.

Concerning general magnetic potentials, HEGNN and SEGNN work in ensembles that are expressed as

$$E^{\text{total}} = E^{\text{HEGNN}} + E^{\text{SEGNN}} \qquad (1)$$

$$E^{\text{HEGNN}}(\vec{r}, \vec{s}) = \lambda_r E_r + \lambda_{HB} E_{HB}$$

$$= \lambda_r \sum_i h_i^{\text{HEGNN}}(\vec{r}) + \lambda_{HB} \sum_{i,j} J_{ij}^{\text{HEGNN}}(r)\vec{s_i} \cdot \vec{s_j} \quad (2)$$

$$E^{\text{SEGNN}}(r,\vec{s}) = \lambda_{rs} \sum_i h_i^{\text{SEGNN}}(r,\vec{s}) \quad (3)$$

where $h_i^{\text{HEGNN}}(r)$ is the HEGNN component of the local atomic energy of atom $i$ and $J_{ij}^{\text{HEGNN}}(r)$ represents the Heisenberg coefficient of the bond between atom $i$ and $j$. Both terms are solely dependent on the neighboring atomic position environment and do not involve the spin configuration. Here, $h_i^{\text{SEGNN}}(r,\vec{s})$ represents the SEGNN component of the local atomic energy of atom $i$ which relates to the position and spin vector of the surrounding atoms. The coefficient, $\lambda_r$, $\lambda_{HB}$ and $\lambda_{rs}$, are used in the potential, as PES is usually dominated by position-related energy, then refined by the Heisenberg interaction and high-order spin-lattice interaction. The spin force, like force and stress, can be calculated using auto-differentiation, as per its definition: the negative gradient of the total energy with respect to the spin vector of atom $i$. It is denoted as $\vec{\omega_i} = -\partial E/\partial \vec{s_i}$ and is used for carrying spin-lattice dynamics. It can also be used for training if provided in the dataset.

The input vector $\vec{s}$ in HEGNN and SEGNN is a unit vector. The length of the magnetic moment can be predicted by a separate MLP in the output of HEGNN or SEGNN if it varies with the atomic structure and spin configuration. When the length of the magnetic moment remains the same across different atomic structures and spin configurations, such as in multiferroic[43,44], it can be excluded from the SpinGNN prediction and treated as a constant during spin-lattice dynamics.

**HEGNN:** We first introduce the HEGNN based on the Heisenberg model and GNN. The Heisenberg model is commonly used to describe magnetic interactions in materials, and it is expressed as $H_{HB} = \sum_{i>j} J_{ij}(r)\vec{s_i} \cdot \vec{s_j}$ where $J_{ij}(r)$ depends on the atomic environment of the bond between two magnetic atoms. However, defining the explicit Heisenberg coefficient $J_{ij}(r)$ is challenging as it is affected by the relative positions of the surrounding atoms, which are difficult to define explicitly. Typically, the Heisenberg coefficient is simplified as $J_{ij}(r_{ij})$ assuming it is only influenced by the atomic distance between two magnetic atoms, thus ignoring the many-body effect. GNNs such as Dimenet++, overcome this limitation by updating the edges of the graph and reflecting the surrounding atomic information of the magnetic bond. As a result, the updated edge features in the line graph can represent the learned Heisenberg coefficient $J_{ij}(r)$ including information about the neighbor atoms of the magnetic bond. The updated node features, on the other hand, represent the local atomic energy, similar to the usual GNN potential. Thus, the magnetic potential of HEGNN can be expressed as Eq. (2), where $E_i^{pos}(r)$ represents the traditional structure potential, derived from the learned node features, and $J_{ij}(r)$ represents the Heisenberg coefficient, derived from the learned edge feature.

**SEGNN:** Then, we introduce the SEGNN model, which utilizes a spin-distance edge crystal graph based on the molecular and crystal GNN with the spin-distance edge feature. Radial basis functions, such as Gaussian expansion[17] and Bessel functions with atomic distance as the only input[18], are typically used for edge initialization of the graph. We propose a spin-distance edge that incorporates bond information on not only the distance but also the spin product, $\vec{s_i} \cdot \vec{s_j}$. SEGNN has two types of edges: magnetic edges and non-magnetic edges. Magnetic edges exist between two magnetic atoms with magnetic moments, whereas non-magnetic edges occur when at least one of the two atoms lacks magnetic moments. To construct the spin edge, we use an expansion basis of the value of $\vec{s_i} \cdot \vec{s_j}$ for magnetic edges, and we use a zero array for non-magnetic edges to distinguish them from the magnetic ones with $\vec{s_i} \cdot \vec{s_j} = 0$. Then, we stack the spin edge and distance edge to form the final spin-distance edge. The implementation of the spin-distance edge could be expressed as

$$e_{ij}^{\text{distance}} = f\left(rbf(|\vec{r_{ij}}|)\right) \tag{4}$$

$$e_{ij}^{\text{spin}} = f\left(Basis(\vec{s_i} \cdot \vec{s_j})\right) \text{ for magnetic edges} \tag{5}$$

$$e_{ij}^{\text{spin}} = f(Basis(0.0) \times 0.0) \text{ for non-magnetic edges} \tag{6}$$

$$e_{ij}^{\text{spin-distance}} = f\left(e_{ij}^{\text{distance}} \oplus e_{ij}^{\text{spin}}\right) \tag{7}$$

where $\oplus$ denotes the concatenation. The addition of spin-distance edges provides a simple and effective way to include spin information in the original crystal graph. Therefore, any GNN model can be used to represent magnetic interactions using this modification in the edge initialization process to construct the magnetic potential.

**Spin-Dimenet++**. Dimenet++[18,19], a remarkable line graph neural network for materials, is applied within SpinGNN and used to construct the magnetic potential shown in Fig. 2a. In HEGNN-DimeNet++, the last layer's edge message features $\boldsymbol{m}_{ji}$ of DimeNet++ are employed to establish the edge Heisenberg coefficients using multilayer perceptrons (MLPs) to map high-dimensional features to a single-value scalar. Meanwhile, in SEGNN-DimeNet++, the spin-distance edge $\boldsymbol{e}_{RBF-New}^{(ji)}$ replaces the distance representations $\boldsymbol{e}_{RBF}^{(ji)}$, an amalgamation of the original $\boldsymbol{e}_{RBF}^{(ji)}$ and the spin dot-product of the two atoms on the bond $\langle \vec{s_i}, \vec{s_j} \rangle$ obtained employing some MLPs with details shown in Fig. 2a.

**Spin-Allegro**. Allegro[23], a strictly local equivariant neural network interatomic potential architecture, is implemented within SpinGNN and utilized to create the magnetic potential and perform large-scale spin-lattice dynamics, shown in Fig. 2b. HEGNN-Allegro and SEGNN-Allegro are stacked within the Spin-Allegro. HEGNN-Allegro constructs the local edge energy along with the edge Heisenberg coefficients, utilizing distinct MLPs, using updated edge features as input. For SEGNN-Allegro, the edge features from HEGNN-Allegro are combined with $\langle \vec{s_i}, \vec{s_j} \rangle$ expansion based on the Fourier basis, the one-hot embedding of atom species, and the spin information of the center and neighbor atom to serve as input features for the next layer of Allegro:

$$\mathbf{x}_{\text{SEGNN}}^{ij, L=0} = \text{MLP}_{\text{two-body}}(1\text{Hot}(Z_i) \oplus 1\text{Hot}(Z_j) \oplus |\vec{s_i}| \oplus |\vec{s_j}| \oplus \mathbf{x}_{\text{HEGNN}}^{ij, L=last} \oplus B(\vec{s_i} \cdot \vec{s_j})) \cdot u(\vec{r_{ij}}) \tag{8}$$

where ⊕ denotes concatenation, $\text{1Hot}(\cdot)$ is a one-hot encoding of the center and neighbor atom species $Z_i$ and $Z_j$, $|\vec{s_i}|$ is the normalization of the spin vector, $\mathbf{x}_{\text{HEGNN}}^{ij,\,L=last}$ is the edge invariant features from the last layer of the HEGNN-Allegro, $B(\vec{s_i}\cdot\vec{s_j})$ is the projection of the $\vec{s_i}\cdot\vec{s_j}$ onto a Fourier basis and $u(\vec{r_{ij}})$ is the polynomial envelope function as proposed in ref. [18]. While the edges that involve non-magnetic atoms use an array of zero to replace the $\langle\vec{s_i},\vec{s_j}\rangle$ expansion $B(\vec{s_i}\cdot\vec{s_j})$, the magnetic edges require both atoms of the edges to be magnetic. The trained Spin-Allegro model can be utilized to perform spin-lattice calculations using the SPIN package in LAMMPS[39,45,46]. Furthermore, it is parallelizable across multiple workers and can efficiently perform large-scale spin-lattice dynamics, similar to the large-scale molecular dynamics achievable with Allegro[23].

**Stress calculation for Allegro in parallel.**

In neural network potential, stress can be calculated by the derivatives of a zero strain on the cell over the energy with the help of auto-grad technology. However, stress calculation with the cell is limited during parallel calculation in large systems as the cell is not usually treated as an input. In LAMMPS[46], a widely used molecular dynamics program, ghost atoms instead of the cell and other periodic boundary information are involved during parallel calculation, which further limits stress calculation with the cell in parallelism. We propose an edge-vector-based atomic virial method for the parallel stress calculation of Allegro[23], which makes it possible to carry out NPT simulations in parallel.

We first calculate the virial $\Xi$ of the system and stress is equal to $\frac{\Xi}{\Omega}$ where $\Omega$ is the volume of the cell. According to the virial definition, $\Xi = \sum_i \boldsymbol{R}_i \boldsymbol{F}_i$ with $\boldsymbol{R}_i$ and $\boldsymbol{F}_i$ as the position and force vector of $i$'s atom. In Allegro, the energy can be expressed as $E = \sum_{i,j} E_{ij}$ where coefficients related to all the species are ignored for clarity. We define node energy as $E_i = \sum_j E_{ij}$ and note that $E_{ij} \neq E_{ji}$ for the edge energy in Allegro. As the atomic energy is strictly local in Allegro and the position information of atoms is collected as $\boldsymbol{R}_{ij}$ with the center atom $i$ and neighbor atom $j$, the edge energy can be expressed as $E_{ij} = f_1(Z_i, \{Z_j, \boldsymbol{R}_{ij}, j \in N(i)\})$ and similarly, node energy as $E_i = f_2(Z_i, \{Z_j, \boldsymbol{R}_{ij}, j \in N(i)\})$. Here, $\boldsymbol{R}_{ij}$ is defined as $\boldsymbol{R}_{ij} = \boldsymbol{R}_j - \boldsymbol{R}_i$ and the system energy is $E = \sum_i E_i$. For atom i as the center atom, we can derive that $\frac{\partial E_i}{\partial \boldsymbol{R}_i} = -\sum_{j \neq i} \frac{\partial E_i}{\partial \boldsymbol{R}_{ij}}$ for atom $i$ as the center atom and $\frac{\partial E_j}{\partial \boldsymbol{R}_i} = \frac{\partial E_j}{\partial \boldsymbol{R}_{ji}}$ for atom $i$ as the neighbor atoms. As a result, the force is expressed as $\boldsymbol{F}_i = -\frac{\partial E}{\partial \boldsymbol{R}_i} = -\frac{\partial \sum_j E_j}{\partial \boldsymbol{R}_i} = -\frac{\partial E_i}{\partial \boldsymbol{R}_i} - \sum_{j \neq i} \frac{\partial E_j}{\partial \boldsymbol{R}_i} = \sum_{j \neq i} \frac{\partial E_i}{\partial \boldsymbol{R}_{ij}} - \sum_{j \neq i} \frac{\partial E_j}{\partial \boldsymbol{R}_{ji}}$. Hence, the virial can be expressed as

$$\Xi = \sum_i \boldsymbol{R}_i \boldsymbol{F}_i = \sum_i \boldsymbol{R}_i \left( \sum_{j \neq i} \frac{\partial E_i}{\partial \boldsymbol{R}_{ij}} - \sum_{j \neq i} \frac{\partial E_j}{\partial \boldsymbol{R}_{ji}} \right)$$

$$= \sum_i \boldsymbol{R}_i \sum_{j \neq i} \frac{\partial E_i}{\partial \boldsymbol{R}_{ij}} - \sum_i \boldsymbol{R}_i \sum_{j \neq i} \frac{\partial E_j}{\partial \boldsymbol{R}_{ji}}$$

$$= \sum_i \boldsymbol{R}_i \sum_{j \neq i} \frac{\partial E_i}{\partial \boldsymbol{R}_{ij}} - \sum_j \boldsymbol{R}_j \sum_{i \neq j} \frac{\partial E_i}{\partial \boldsymbol{R}_{ij}}$$

$$= -\sum_i \sum_{j \neq i} \boldsymbol{R}_{ij} \frac{\partial E_i}{\partial \boldsymbol{R}_{ij}} \tag{9}$$

Considering the symmetry of the virial and the numerical error in the calculation, the final virial is calculated with $\Xi_i = -\sum_{j \neq i} (\boldsymbol{R}_{ij} \frac{\partial E_i}{\partial \boldsymbol{R}_{ij}} + \boldsymbol{R}_{ji} \frac{\partial E_j}{\partial \boldsymbol{R}_{ji}})/2$ as the atomic virial for each atom and $\Xi = \sum_i (\Xi_i + \Xi_i^T)/2$ as the system virial. $\Xi_i$ can be calculated with only center and neighbor atoms' positions in parallel and $\frac{\partial E_i}{\partial \boldsymbol{R}_{ij}}$ can be easily calculated with the auto-grad technique. This method requires only the positions of the atoms as input without the cell, and the virial of the entire system can be calculated by the accumulation of atomic virial that can be correctly calculated in parallel. Additionally, the atomic virial provided in this method helps obtain the correct heat current for simulations of heat transport [47,48]. A careful numerical check is carried out, ensuring consistency with the results of the zero-strain cell stress method and LAMMPS parallel computation. With this edge-vector-based atomic virial method, we can carry out NPT simulations in parallel and simulate heat transport with Allegro. Implementations are available at https://github.com/Hongyu-yu/nequip and https://github.com/Hongyu-yu/pair_allegro.

**Fitting the artificially constructed spin model**

SpinGNN's capability to precisely learn energies from different artificially constructed spin models is demonstrated with Spin-DimeNet++. As depicted in Fig. 2a, Spin-DimeNet++ precisely models the intricate spin-spin and spin-lattice interactions, achieving high accuracy.

***Spin Hamiltonian of an artificial model***. We first validate Spin-DimeNet++ by developing a spin Hamiltonian with a fixed structure that only accounts for spin degrees of freedom. We utilize an artificial model that includes Heisenberg interactions, 4th-order biquadratic interactions, and 4-body 4th-order interactions to describe the spin interactions between magnetic ions occupying the Al sites of the CuAlO2-type[49] structure. The spin Hamiltonian model consists of intralayer interactions within a triangular lattice and interlayer couplings:

$$H = \sum_{<ij>_n}^{n=1,\ldots,5} J_n \mathbf{S}_i \cdot \mathbf{S}_j + \sum_{<ij>_n}^{n=1,5,7} K_n \left(\mathbf{S}_i \cdot \mathbf{S}_j\right)^2 + \sum_{<ijkl>_1} L_1 \left(\mathbf{S}_i \cdot \mathbf{S}_j\right)\left(\mathbf{S}_k \cdot \mathbf{S}_l\right). \#(10)$$

For more details regarding this model, please refer to [50]. Since high-order interactions must be considered, HEGNN and SEGNN are used to construct the GNN spin Hamiltonian, fitting the model well, as shown in Fig. 3a, with a mean absolute error (MAE, defined as $\frac{\sum |f(x_i)-y_i|}{n}$) of $1.024 \times 10^{-5}$ meV/site. This performance is slightly better than that of the machine learning spin Hamiltonian based on the spin descriptor, with an MAE of $1.03 \times 10^{-5}$ meV/site[50]. Thus, Spin-DimeNet++ can effectively reconstruct the energy landscape of a complex spin Hamiltonian.

***Two artificial spin-lattice models***. We assess the capability of Spin-DimeNet++

to develop a magnetic potential that considers both structural and magnetic degrees of freedom. To fully validate Spin-DimeNet++, we used two complex spin-lattice model potentials that include different types of magnetic interactions. The first model potential is a Heisenberg-type, while the second one includes higher-order magnetic interactions. These model potentials are meant to describe the spin-lattice couplings in the magnetic B-sites of the $ABO_3$ perovskite structure and embody a total of 50 atoms.

The first model potential considered does not include high-order spin interactions. Instead, it consists of 108 terms expressed below, encompassing up to three-body and second-order interactions that are expressed as

$$H = \sum_{<ijk>_r}^{r<7\text{Å}} J_{33}(r_i, r_j, r_k) u_{i/j/k} (\mathbf{S}_i \cdot \mathbf{S}_j) + J_{22}(r_i, r_j, r_k)(\mathbf{S}_i \cdot \mathbf{S}_j) + E_3(r_i, r_j, r_k) + \sum_{<ij>_r}^{r<7\text{Å}} J_{23}(r_i, r_j) u_{i/j} (\mathbf{S}_i \cdot \mathbf{S}_j) + J_{22}(r_i, r_j)(\mathbf{S}_i \cdot \mathbf{S}_j) + E_2(r_i, r_j) \qquad (11)$$

where $J_{ij}$ indicates the Heisenberg coefficient of i-body and $j^{th}$-order which is all set as 0.1 eV, while all other coefficients of terms only related to positions are set to 1 eV. The spin-lattice model potential used in this study is a complex Heisenberg model that does not consider high-order magnetic interactions. Since the first model only involves Heisenberg interactions, we have utilized only HEGNN-DimeNet++ for training the model. The results, as shown in Fig. 3a, are impressive, with a mean absolute error (MAE) of 5.04 meV/site. We have also attempted to improve the results by ensembling HEGNN-DimeNet++ and SEGNN-DimeNet++. However, this did not produce a better outcome, and the MAE was 5.57 meV/site due to the overfitting of SEGNN-DimeNet++. As a result, we conclude that HEGNN-DimeNet++ is efficient enough to learn the Heisenberg-type magnetic potential, indicating its ability to grasp the Heisenberg coefficients accurately.

The second model potential introduces fourth-order spin interactions, consisting of 274 terms, encompassing up to three-body and fourth-order interactions that are expressed as

$$H = \sum_{<ijk>_r}^{r<5.71\text{Å}} J_{34}(r_i,r_j,r_k)u_{i/j/k}u_{i/j/k}(\mathbf{S}_i \cdot \mathbf{S}_j) + J_{33}(r_i,r_j,r_k)u_{i/j/k}(\mathbf{S}_i \cdot \mathbf{S}_j)J_{32}(r_i,r_j,r_k)(\mathbf{S}_i \cdot \mathbf{S}_j)$$
$$+ L(r_i,r_j,r_k)(\mathbf{S}_i \cdot \mathbf{S}_j)(\mathbf{S}_i \cdot \mathbf{S}_k) + E_3(r_i,r_j,r_k) +$$
$$\sum_{<ij>_r}^{r<5.71\text{Å}} J_{24}(r_i,r_j)u_{i/j}u_{i/j}(\mathbf{S}_i \cdot \mathbf{S}_j) + J_{23}(r_i,r_j)u_{i/j}(\mathbf{S}_i \cdot \mathbf{S}_j) + J_{22}(r_i,r_j)(\mathbf{S}_i \cdot \mathbf{S}_j) + E_2(r_i,r_j) \quad (12),$$

where coefficients are set the same as the first model that spin-related coefficients at 0.1 eV and all others at 1 eV. The model potential comprises both Heisenberg-type terms and fourth-order terms $L(r_i,r_j,r_k)(\mathbf{S}_i \cdot \mathbf{S}_j)(\mathbf{S}_i \cdot \mathbf{S}_k)$. The fourth-order terms cannot be learned by HEGNN-DimeNet++ alone. Initially, only HEGNN-DimeNet++ was used for fitting, resulting in a high MAE of 2.26 meV/atom in tests as shown in Fig. 3b. Due to the high-order interactions in the second model, HEGNN-DimeNet++ alone is insufficient, and SEGNN-DimeNet++ is required. Then, only SEGNN-DimeNet++ was used for fitting, resulting in a relatively good MAE of 0.596 meV/atom as shown in Fig. 3c. Subsequently, HEGNN and SEGNN were used jointly for fitting, resulting in a better MAE of 0.466 meV/atom as shown in Fig. 3d. SEGNN-DimeNet++ effectively learns high-order spin-lattice interactions, while HEGNN-DimeNet++ excels in learning Heisenberg interactions, and the combination of the two models improves the overall accuracy.

**Application of Multiferroic BiFeO$_3$**

We utilized SpinGNN to explore its capability in representing magnetic materials featuring large-scale spin-lattice dynamics. Specifically, we employed Spin-Allegro to investigate BiFeO$_3$ (BFO), the room-temperature multiferroic materials[33,34,51–56]. BFO has garnered significant interest due to its exclusive combination of ferroelectric and antiferromagnetic properties. BFO possesses a perovskite crystal structure, and its magnetic ground state is antiferromagnetic. BFO displays a phase transition temperature around 643 K[56], identified as antiferromagnetic Néel temperature, that results from the interplay of different magnetic interactions. Comprehending the mechanism behind the phase transitions in BFO is crucial to design and optimizing materials for a wide range of applications, including but not limited to data storage, spintronics, and energy harvesting.

A BFO dataset comprising 4857 data points is generated through first-principles calculations. Spin-Allegro, comprising two layers of HEGNN-Allegro and one layer of SEGNN-Allegro, achieves high accuracy on the test dataset, with an MAE of 1.40 meV/site and 23.2 meV/Å for energy and force accuracy, respectively. In contrast, the original Allegro with the same configuration is trained with an MAE of 3.55 meV/site and 30.1 meV/Å for energy and force accuracy, respectively.

To validate the potential accuracy, we utilized the potential along with the conjugate gradient (CG) algorithm[57] to explore the magnetic ground state of BFO. The investigation began with a 12×12×12 supercell composed of 8640 atoms and 1728 Fe atoms with random spin configurations. The subsequent 1000 CG iterations produced a zero spin temperature perfect G-type antiferromagnetic state, which agrees with the experimental data[51–53]. A small cell of BFO structure with a G-type antiferromagnetic state is shown in Fig. 4a.

Then we utilized spin-lattice SIB-NPT simulation[44,58] over a range of temperatures to examine the magnetic phase transition of BFO, starting from the same structure as earlier. We collected the antiferromagnetic vector from 50K to 1200K, as shown in Fig. 4b. As the temperature increased, the antiferromagnetic vector decreased, indicating a magnetic transition from the antiferromagnetic to the paramagnetic state. The Néel temperature was estimated to be 650K, consistent with experiments[51–53]. We conducted a spin dynamics simulation without atomic motions and confirmed that spin-lattice coupling drives a lower and more accurate transition temperature shown in Fig. 4b. The Spin-Allegro model provides a precise depiction of various atomic and spin phenomena at different temperatures. These results indicate that the model effectively characterizes the intricate spin-lattice interaction of BFO.

Furthermore, we investigate the domain wall (DW) of $BiFeO_3$ (BFO), which plays a crucial role in its properties and applications. Different DW types in BFO have been observed through experimental studies, such as DW109, DW180, and DW71, with characteristic angles of 109°, 180°, and 71°, respectively. These DWs can be controlled by external conditions such as magnetic fields, temperature, and pressure, and have gained attention for their potential applications[59–61]. However, the complex and large structure of the DWs in BFO presents challenges for effectively optimizing their spin orientation and configuration and understanding their various properties[62–66]. To address this, we optimize several DW structures using the CG and annealing algorithm[67] with Spin-Allegro and investigate their meta-stable structures and related energy. We analyzed the DW energies of DW109, DW180, and DW71 by calculating $(E_{DW} - E_{ferro})/2/A$, where A is the area of the supercell's topmost layer and $E_{ferro}$ is the energy of pure ferroelectric structure. Initially, we created relatively small cell DWs with 160, 80, and 80 atoms for DW109, DW180, and DW71, respectively. The DW energies of these DWs are in the order of 65.70 J/m2, 166.4 J/m2, and 182.0 J/m2, respectively, which is in line with previous studies and presents a relatively consistent value. The optimized structures were analyzed using DFT, and the forces on the atoms are minimal, demonstrating the efficacy of optimization with Spin-Allegro. Subsequently, we constructed larger DW structures with 1280, 640, and 640 atoms for DW109, DW180, and DW71, respectively, by increasing the size of the small DW structures' supercell to 2 × 2 × 2. The DW energies of these structures are found to be 65.27 J/m$^2$, 170.5 J/m$^2$, and 184.6 J/m$^2$, respectively, which are relatively close to those of the small structures. All simulations ended with an antiferromagnetic state. Detailed energy values of the structures and workflow of optimization are provided in SM.

## Discussion

We introduce a novel deep-learning magnetic interatomic potential suitable for magnetic materials that account for the freedom degrees of atomic position and noncollinear magnetic moment. We achieve high accuracy in this endeavor by proposing a Spin Graph Neural Network. Our resulting Spin-DimeNet++ and Spin-Allegro models, based on the SpinGNN framework, exhibit good performance on the magnetic dataset, furnishing a dependable energy landscape for an artificial high-order spin Hamiltonian, two artificial complex spin-lattice models, as well as the multiferroic BiFeO$_3$. Spin-Allegro is equipped to scale to large system sizes with ease thanks to the strict locality of its geometric representations in Allegro, along with parallelization in large-scale calculations. This renders it feasible to examine the magnetic, structural, and kinetic properties of complex magnetic systems that comprise millions of atoms with spin-lattice dynamics simulations at nearly first-principles fidelity. Given that the magnetic potential is important for simulations of magnetic materials and understanding magnetism, we expect that our framework will find wide use in the study of magnetism.

Our framework facilitates the study of previously inaccessible magnetic materials systems with graph neural networks. It is worth noting that the SpinGNN framework can be implemented with most graph neural network architectures, such as MACE[25], GemNet[20], NequIP[22], and ALIGNN[24] to be implemented just as DimeNet++[19] and Allegro[23]. Compared to previous works that built magnetic potential based on different descriptors with different implementations, SpinGNN serves as a general framework to transfer the common neural network potential into a magnetic potential. Furthermore, just as the GNN potential usually outperforms potential based on descriptors[68], the magnetic GNN potential with the SpinGNN framework is expected to be more powerful than the previously proposed magnetic potential based on descriptors, although a benchmark magnetic dataset is not yet available. With the continuous development of the neural network potential, particularly in recent years, we expect a better magnetic neural network potential to be built with a more powerful graph neural network potential and SpinGNN framework. Spin-Allegro is based on one of the most popular graph neural network potentials and has been proven to be not only powerful enough but also easy to parallelize for large-scale spin-lattice dynamics. Also, the time-reversal symmetry is naturally conserved in SpinGNN, where the feature of the spin-distance edge is about $\vec{s_i} \cdot \vec{s_j} = (-\vec{s_i}) \cdot (-\vec{s_j})$ and the Heisenberg-type spin-lattice formula $J_{ij}^{\text{HEGNN}}(r)\vec{s_i} \cdot \vec{s_j} = J_{ij}^{\text{HEGNN}}(r)(-\vec{s_i}) \cdot (-\vec{s_j})$.

## Methods

### Software.

We developed our codes utilizing the modified NequIP code available at https://github.com/Hongyu-yu/nequip under git commit bb10cc6f91ca42fda452d7b22cc314d33c0e3d95, Allegro code available at https://github.com/mir-group/allegro under git commit a7899a9c4c6be0620e11bef0bca8d06d9f7f32a8, as well as e3nn[69] with version 0.5.0, Pytorch[70] with version 1.11.0 and Python with 3.9.13. Spin-DimeNet++ is developed based on the implementation in Pytorch_geometric[71], which is available at https://github.com/pyg-team/pytorch_geometric/blob/2.0.4/torch_geometric/nn/models/dimenet.py. The LAMMPS was built based on the LAMMPS code available at https://github.com/lammps/lammps.git under git commit 9b989b186026c6fe9da354c79cc9b4e152ab03af with the modified pair_allegro code available at https://github.com/Hongyu-yu/pair_allegro, git commit 101d2c8db123c9549490fa7285ce5d85a616d46f. The VESTA[72] software was used to generate Fig. 1 and Fig. 4a. Matplotlib[73] was used for plotting results.

**Spin-lattice dynamics simulations.**

We performed spin-lattice dynamics simulations using the SPIN package [39] in LAMMPS[46]. To ensure the accurate integration of the Landau-Lifshitz-Gilbert (LLG)[74] equation and enforce the conservation of the magnetic moments' magnitude[44], we implemented the semi-implicit SIB method introduced by Mentink et al.[58] in LAMMPS and the SPIN package. The dynamics of spins and atomic motions were simulated using the molecular dynamics scheme (NVT or NPT) and the SIB method. Our simulations of BFO were carried out on a periodic 12×12×12 supercell comprising 8640 atoms with a time step of 0.5 fs across 4 NVIDIA A100 GPUs simultaneously in NPT-SIB. Temperature control was exerted by implementing the Nosé Hoover thermostat with a temperature damping parameter of 100 time steps, and pressure control was maintained using the Parrinello-Rahman barostat with a temperature damping parameter of 1000 time steps, as implemented in LAMMPS[45]. To model the temperature-driven antiferromagnetic-paramagnetic, we simulated the dynamics of atomic motions and spin motions using NPT simulations, respectively, with a damping coefficient in the LLG equation of 0.01. The simulation commenced with BFO in the *R3c* phase, in which the antiferromagnetic configurations were initialized randomly in 100 K with a Gaussian distribution of velocity. The system was heated from 100 K to 1200 K in 30 ps and then cooled from 1200 K to 50 K in 120 ps in steps of 50 K. At each given temperature, an equilibrium run of 5 ps was followed by a production run of 10 ps.

**Reference training sets.**

We build the dataset of artificially constructed spin models using the effective Hamiltonian in the Property Analysis and Simulation Package (PASP)[35]. The first dataset of spin Hamiltonian is established on $CuMO_2$, where M is a magnetic ion, with a fixed $CuAlO_2$-type structure. We generate a dataset of 10,000 fully randomized spin configurations based on a 4×4×2 supercell with 96 magnetic sites, i.e., 6 layers of 4×4 triangular lattices, and obtain their corresponding energies from Eq. (10). We refer the reader to our previous work[50] for more details. For the two spin-lattice effective Hamiltonian datasets, we construct atomic configurations based on the ideal structures of $ABO_3$ perovskite with 50 atoms in total, comprising magnetic B sites. We randomly sample spin configurations and calculate energy based on the constructed effective Hamiltonian from Eq. (11) and Eq. (12). In total, we generate 2416 and 3611 data with different distortions and spin configurations for the two spin-lattice Hamiltonians, respectively.

For the BFO dataset, we perform Density Functional Theory (DFT) calculations using VASP[75–77] with the Projector Augmented Wave (PAW) method[78] and the Perdew–Burke–Ernzerhof (PBE) functional[79]. We generate atomic configurations through several NPT molecular dynamics simulations starting from 4×2×2 supercells with 80 atoms under different phases, including *R3c*, *R3m*, *Pnma*, *and Pm-3m*. We also randomly sample spin configurations in the calculations without considering spin-orbit couplings[80]. We choose an energy cutoff for the plane wave basis of 500 eV and a 2×4×4 k-point grid. Total energy calculations are converged to $10^{-6}$ eV per supercell to ensure high-accuracy DFT data. We use a self-consistent value of the effective Hubbard U[81] of 3.8 eV for the localized 3d electrons of Fe ions[82].

**Training details.**

Models were trained on an NVIDIA A100 GPU in single-GPU training with the Mean Squared Error (MSE) loss function and the Adam optimizer[83] in Pytorch[70] with default parameters of $β_1$ =0.9, $β_2$ = 0.999, and $ε$ =$10^{-8}$. We lowered the learning rate with an on-plateau scheduler that responded to the validation loss. The learning stopped when the learning rate dropped below $10^{-6}$ or no advancement was made in the validation loss for multiple epochs. All models were trained with float32 precision. We utilized the SiLU activation function [84].

**Spin Hamiltonian and Spin-lattice Hamiltonians.** Spin-DimeNet++ was implemented with identical parameters across all three datasets. The model was trained on energy prediction as the target variable, with 4 layers, each consisting of 128 hidden channels. In addition, we employed 64 interaction triplet embeddings, 8 basis transformation embeddings, 256 output embedding channels, 7 spherical harmonics basis, and 5 for envelope exponent. Three linear layers were used for the output blocks, and both interaction blocks before and after the skip connection were composed of 1 and 2 residual layers, respectively. The batch sizes were 10, 6, and 8 for the spin Hamiltonian, Heisenberg-type, and high-order spin-lattice Hamiltonian, respectively. We initialized the learning rate to be 0.0005 for all models, with a learning rate patience of 5, a decay factor of 0.8, and a stop patience of 20. We chose cutoff values of 8 Å, 7.2 Å, and 6 Å based on the Hamiltonian construction to ensure sufficient information was included. The Adam optimizer's weight decay was 0.01. Finally, we randomly partitioned the data into three sets: 80% for training, 10% for validation, and 10% for testing.

**$BiFeO_3$.** We split our data into training (4200 samples), validation (357 samples), and testing (300 samples) sets. The target variables were energies, forces, and stresses. To define the loss function, we used per-atom mean squared error[23], with weights of 1 assigned to the energy, force, and stress terms, and no weight decay in the Adam optimizer. For HBGNN-Allegro, we employed two layers, 16 features for both even and odd irreps, and a $\ell$max = 2. We used one layer for SEGNN-Allegro. The two-body latent MLP for both models consisted of three hidden layers with dimensions [64, 128, 128] using SiLU nonlinearities. The later latent MLPs consisted of one hidden layer of dimension 128 using SiLU nonlinearity. We implemented the embedding weight projection using a single matrix multiplication. For HEGNN-Allegro, the final edge energy and Heisenberg coefficient MLPs both had three hidden layers with dimensions [128, 128, 64], and SiLU nonlinearities, while for SEGNN-Allegro, the final edge energy MLP had one hidden layer of dimension 128 without nonlinearity. The weight initialization for all MLPs used a uniform distribution of unit variance. The coefficients $\lambda_r$, $\lambda_{HB}$, and $\lambda_{rs}$ in Eq. 1-3 and Fig. 2 were set to 1.0, 0.05, and 0.05, respectively. We set the radial cutoff at 6.0 Å and utilized an 8 non-trainable Bessel function basis for encoding with the polynomial envelope function using p=48. The learning rate was set to 0.005, batch size to 1, with an on-plateau scheduler reducing the learning rate based on the validation loss of force every time it hit a plateau. The patience was set to 5, and the decay factor was set to 0.3. An exponential moving average with a weight of 0.99 was used to evaluate the validation set and the final model. We stopped the training when the learning rate dropped below 1e-6.

## Data availability

The authors declare that all data supporting the findings of this study are available from the corresponding author on reasonable request.

## Code availability

The stress implementations for NequIP, Allegro, and LAMMPS interface are available at https://github.com/Hongyu-yu/nequip and [https://github.com/Hongyu-yu/pair_allegro](https://github.com/Hongyu-yu/pair_allegro). Additional codes related to this work may be requested from the authors.

\* [hxiang@fudan.edu.cn](mailto:hxiang@fudan.edu.cn)

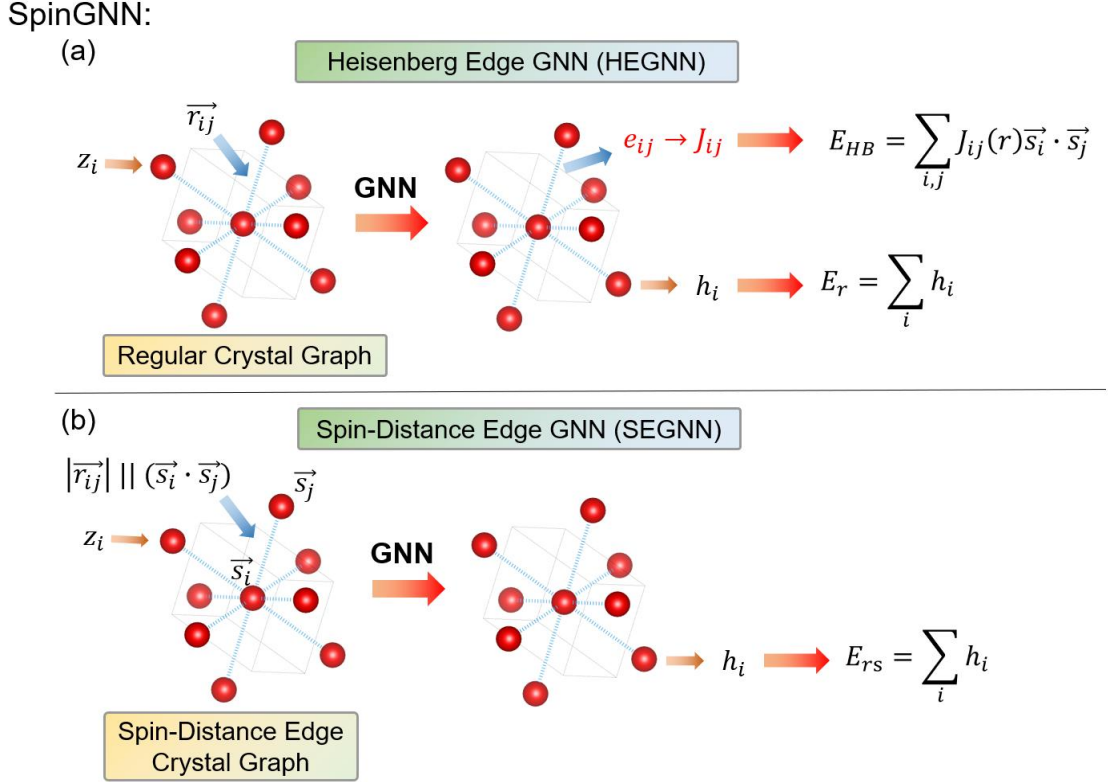

**Figure 1 The SpinGNN framework.** Illustration of the SpinGNN including the Heisenberg Edge GNN (HEGNN) and Spin-Distance Edge GNN (SEGNN). **a** HEGNN utilizes the updated edge feature of GNN as the Heisenberg coefficients and builds a Heisenberg-based magnetic potential. **b** SEGNN utilizes the spin-distance edge crystal graph which initializes the edge with the dot product of the spin vector and bond length and builds a high-order general magnetic potential.

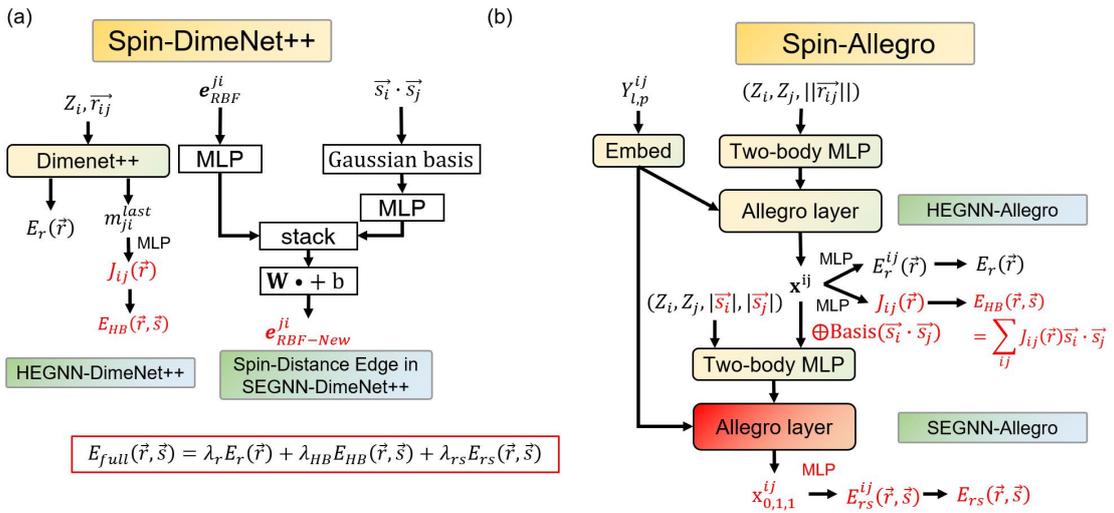

**Figure 2 The Spin-Dimenet++ and Spin-Allegro architecture. a** shows details of the HEGNN-DimeNet++ and spin-distance edge in SEGNN-DimeNet and **b** show details of the Spin-Allegro which is stacked by HEGNN-Allegro and SEGNN-Allegro. Spin-related parts are highlighted in red.

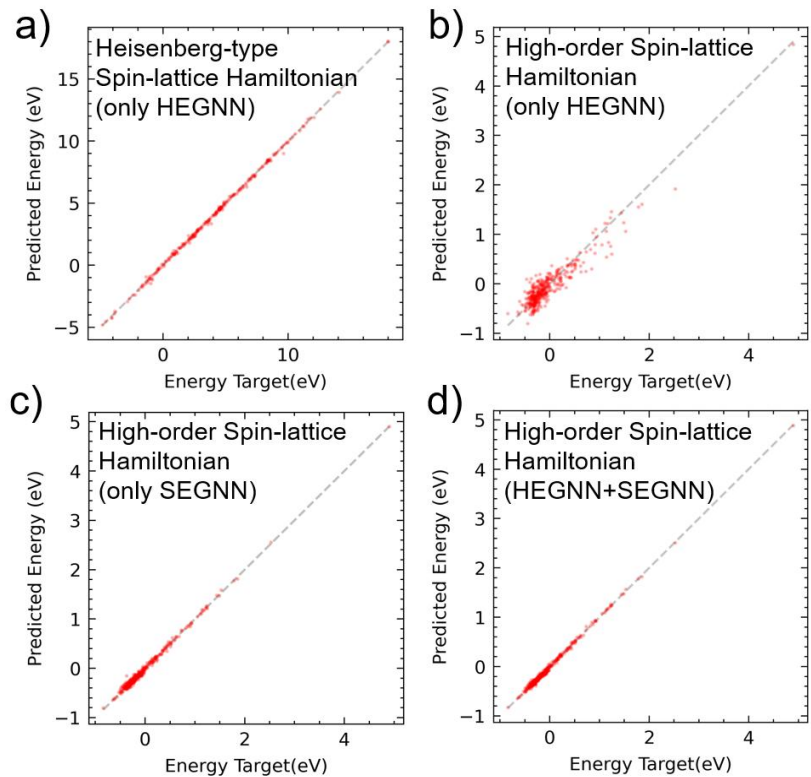

**Figure 3.** The prediction results of Spin-DimeNet++ on the energy of test datasets generated from the artificially constructed spin model.

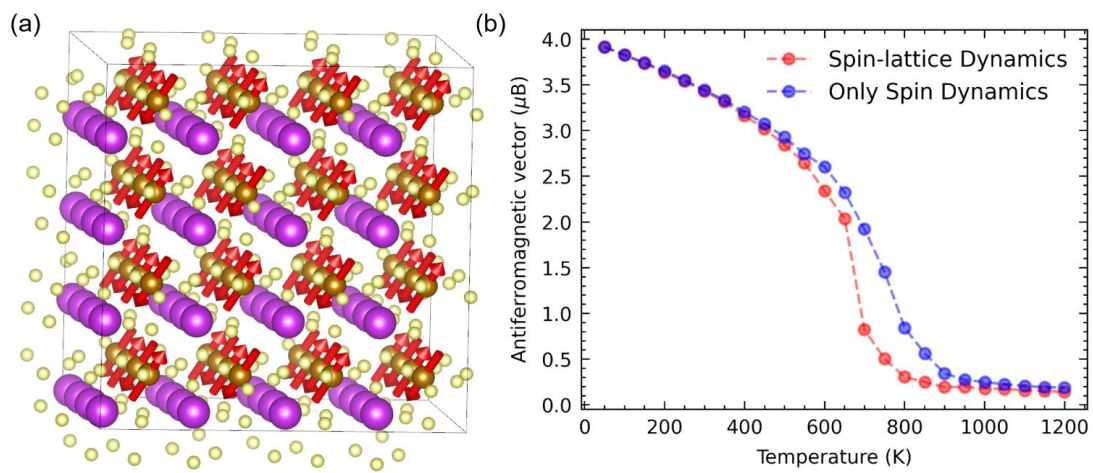

**Figure 4 a** The structure of BiFeO$_3$ with ground G-type antiferromagnetic state. **b** The antiferromagnetic vector in different temperatures of BiFeO$_3$ from simulations of spin-lattice dynamics and spin dynamics with Spin-Allegro. Obvious magnetic transition is observed and transition temperatures can be estimated.

# Reference

1. Carleo, G. *et al.* Machine learning and the physical sciences. *Rev. Mod. Phys.* **91**, 045002 (2019).
2. Agrawal, A. & Choudhary, A. Perspective: Materials informatics and big data: Realization of the "fourth paradigm" of science in materials science. *APL Mater.* **4**, 053208 (2016).
3. Li, X., Li, Z. & Chen, J. Ab initio calculation of real solids via neural network ansatz. *Nat. Commun.* **13**, 7895 (2022).
4. Deringer, V. L. *et al.* Origins of structural and electronic transitions in disordered silicon. *Nature* **589**, 59–64 (2021).
5. Chmiela, S., Sauceda, H. E., Müller, K.-R. & Tkatchenko, A. Towards exact molecular dynamics simulations with machine-learned force fields. *Nat. Commun.* **9**, 3887 (2018).
6. Ko, T. W., Finkler, J. A., Goedecker, S. & Behler, J. A fourth-generation high-dimensional neural network potential with accurate electrostatics including non-local charge transfer. *Nat. Commun.* **12**, 398 (2021).
7. Friederich, P., Häse, F., Proppe, J. & Aspuru-Guzik, A. Machine-learned potentials for next-generation matter simulations. *Nat. Mater.* **20**, 750–761 (2021).
8. Behler, J. & Parrinello, M. Generalized Neural-Network Representation of High-Dimensional Potential-Energy Surfaces. *Phys. Rev. Lett.* **98**, 146401 (2007).
9. Bartók, A. P., Payne, M. C., Kondor, R. & Csányi, G. Gaussian Approximation Potentials: The Accuracy of Quantum Mechanics, without the Electrons. *Phys. Rev. Lett.* **104**, 136403 (2010).
10. Thompson, A. P., Swiler, L. P., Trott, C. R., Foiles, S. M. & Tucker, G. J. Spectral neighbor analysis method for automated generation of quantum-accurate interatomic potentials. *J. Comput. Phys.* **285**, 316–330 (2015).
11. Shapeev, A. V. Moment Tensor Potentials: A Class of Systematically Improvable Interatomic Potentials. *Multiscale Model. Simul.* **14**, 1153–1173 (2016).
12. Zhang, L., Han, J., Wang, H., Car, R. & E, W. Deep Potential Molecular Dynamics: A Scalable Model with the Accuracy of Quantum Mechanics. *Phys. Rev. Lett.* **120**, 143001 (2018).
13. Drautz, R. Atomic cluster expansion for accurate and transferable interatomic potentials. *Phys. Rev. B* **99**, 014104 (2019).
14. Bartók, A. P., Kondor, R. & Csányi, G. On representing chemical environments. *Phys. Rev. B* **87**, 184115 (2013).
15. Behler, J. Atom-centered symmetry functions for constructing high-dimensional neural network potentials. *J. Chem. Phys.* **134**, 074106 (2011).
16. Schütt, K. T., Sauceda, H. E., Kindermans, P.-J., Tkatchenko, A. & Müller, K.-R. SchNet – A deep learning architecture for molecules and materials. *J. Chem. Phys.* **148**, 241722 (2018).
17. Xie, T. & Grossman, J. C. Crystal Graph Convolutional Neural Networks for an Accurate and Interpretable Prediction of Material Properties. *Phys. Rev. Lett.* **120**, 145301 (2018).
18. Klicpera, J., Groß, J. & Günnemann, S. Directional Message Passing for Molecular Graphs.

*ArXiv200303123 Phys. Stat* (2020).

19. Klicpera, J., Giri, S., Margraf, J. T. & Günnemann, S. Fast and Uncertainty-Aware Directional Message Passing for Non-Equilibrium Molecules. *ArXiv201114115 Phys.* (2020).
20. Klicpera, J., Becker, F. & Günnemann, S. GemNet: Universal Directional Graph Neural Networks for Molecules. *ArXiv210608903 Phys. Stat* (2021).
21. Schmidt, J., Pettersson, L., Verdozzi, C., Botti, S. & Marques, M. A. L. Crystal graph attention networks for the prediction of stable materials. *Sci. Adv.* **7**, eabi7948 (2021).
22. Batzner, S. *et al.* E(3)-equivariant graph neural networks for data-efficient and accurate interatomic potentials. *Nat. Commun.* **13**, 2453 (2022).
23. Musaelian, A. *et al.* Learning local equivariant representations for large-scale atomistic dynamics. *Nat. Commun.* **14**, 579 (2023).
24. Choudhary, K. & DeCost, B. Atomistic Line Graph Neural Network for improved materials property predictions. *Npj Comput. Mater.* **7**, 1–8 (2021).
25. Batatia, I., Kovács, D. P., Simm, G. N. C., Ortner, C. & Csányi, G. MACE: Higher Order Equivariant Message Passing Neural Networks for Fast and Accurate Force Fields. Preprint at https://doi.org/10.48550/arXiv.2206.07697 (2023).
26. Schütt, K. T., Unke, O. T. & Gastegger, M. Equivariant message passing for the prediction of tensorial properties and molecular spectra. *ArXiv210203150 Phys.* (2021).
27. Qiao, Z. *et al.* UNiTE: Unitary N-body Tensor Equivariant Network with Applications to Quantum Chemistry. *ArXiv Prepr. ArXiv210514655* (2021).
28. Haghighatlari, M. *et al.* NewtonNet: a Newtonian message passing network for deep learning of interatomic potentials and forces. *Digit. Discov.* **1**, 333–343 (2022).
29. Thölke, P. & De Fabritiis, G. TorchMD-NET: Equivariant Transformers for Neural Network based Molecular Potentials. Preprint at https://doi.org/10.48550/arXiv.2202.02541 (2022).
30. Brandstetter, J., Hesselink, R., van der Pol, E., Bekkers, E. J. & Welling, M. Geometric and Physical Quantities Improve E(3) Equivariant Message Passing. Preprint at https://doi.org/10.48550/arXiv.2110.02905 (2022).
31. Wang, Q. & Zhang, L. Inverse design of glass structure with deep graph neural networks. *ArXiv210406632 Cond-Mat* (2021).
32. Xie, T., Fu, X., Ganea, O.-E., Barzilay, R. & Jaakkola, T. Crystal Diffusion Variational Autoencoder for Periodic Material Generation. *ArXiv211006197 Cond-Mat Physicsphysics* (2021).
33. Xu, C., Xu, B., Dupé, B. & Bellaiche, L. Magnetic interactions in $BiFeO_3$: A first-principles study. *Phys. Rev. B* **99**, 104420 (2019).
34. Xu, B., Dupé, B., Xu, C., Xiang, H. & Bellaiche, L. Revisiting spin cycloids in multiferroic $BiFeO_3$. *Phys. Rev. B* **98**, 184420 (2018).
35. Lou, F. *et al.* PASP: Property analysis and simulation package for materials. *J. Chem. Phys.* **154**, 114103 (2021).
36. Li, X. *et al.* Spin Hamiltonians in Magnets: Theories and Computations. *Molecules* **26**, 803 (2021).
37. Novikov, I., Grabowski, B., Körmann, F. & Shapeev, A. Magnetic Moment Tensor Potentials for collinear spin-polarized materials reproduce different magnetic states of bcc Fe. *Npj Comput. Mater.* **8**, 13 (2022).
38. Eckhoff, M. & Behler, J. High-dimensional neural network potentials for magnetic systems using spin-dependent atom-centered symmetry functions. *Npj Comput. Mater.* **7**, 170 (2021).
39. Tranchida, J., Plimpton, S. J., Thibaudeau, P. & Thompson, A. P. Massively parallel symplectic algorithm for coupled magnetic spin dynamics and molecular dynamics. *J. Comput. Phys.* **372**, 406–425 (2018).
40. Hukushima, K. & Nemoto, K. Exchange Monte Carlo Method and Application to Spin Glass Simulations. *J. Phys. Soc. Jpn.* **65**, 1604–1608 (1996).
41. Nikolov, S. *et al.* Data-driven magneto-elastic predictions with scalable classical spin-lattice dynamics. *Npj Comput. Mater.* **7**, 153 (2021).
42. Parsaeifard, B. & Goedecker, S. Manifolds of quasi-constant SOAP and ACSF fingerprints and the resulting failure to machine learn four-body interactions. *J. Chem. Phys.* **156**, 034302 (2022).
43. Neaton, J. B., Ederer, C., Waghmare, U. V., Spaldin, N. A. & Rabe, K. M. First-principles study of spontaneous polarization in multiferroic $\mathrm{Bi}\mathrm{Fe}\mathrm{O}_{3}$. *Phys. Rev. B* **71**, 014113 (2005).
44. Wang, D., Weerasinghe, J. & Bellaiche, L. Atomistic Molecular Dynamic Simulations of Multiferroics. *Phys. Rev. Lett.* **109**, 067203 (2012).
45. Plimpton, S. Fast Parallel Algorithms for Short-Range Molecular Dynamics. *J. Comput. Phys.* **117**, 1–19 (1995).
46. Thompson, A. P. *et al.* LAMMPS - a flexible simulation tool for particle-based materials modeling at the atomic, meso, and continuum scales. *Comput. Phys. Commun.* **271**, 108171 (2022).
47. Fan, Z. *et al.* Force and heat current formulas for many-body potentials in molecular dynamics simulations with applications to thermal conductivity calculations. *Phys. Rev. B* **92**, 094301 (2015).
48. Gabourie, A. J., Fan, Z., Ala-Nissila, T. & Pop, E. Spectral decomposition of thermal conductivity: Comparing velocity decomposition methods in homogeneous molecular dynamics simulations. *Phys. Rev. B* **103**, 205421 (2021).
49. Ishiguro, T., Kitazawa, A., Mizutani, N. & Kato, M. Single-crystal growth and crystal structure refinement of CuAlO2. *J. Solid State Chem.* **40**, 170–174 (1981).
50. Yu, H. *et al.* Complex spin Hamiltonian represented by an artificial neural network. *Phys. Rev. B* **105**, 174422 (2022).
51. Blaauw, C. & Woude, F. van der. Magnetic and structural properties of BiFeO3. *J. Phys. C Solid State Phys.* **6**, 1422 (1973).
52. Fischer, P., Polomska, M., Sosnowska, I. & Szymanski, M. Temperature dependence of the crystal and magnetic structures of BiFeO3. *J. Phys. C Solid State Phys.* **13**, 1931 (1980).
53. Karpinsky, D. V. *et al.* Thermodynamic potential and phase diagram for multiferroic bismuth ferrite (BiFeO 3 ). *Npj Comput. Mater.* **3**, 1–10 (2017).
54. Selbach, S. M., Tybell, T., Einarsrud, M.-A. & Grande, T. The Ferroic Phase Transitions of BiFeO3. *Adv. Mater.* **20**, 3692–3696 (2008).
55. Song, Y., Xu, B. & Nan, C.-W. Lattice and spin dynamics in multiferroic BiFeO3 and

RMnO3. *Natl. Sci. Rev.* **6**, 642–652 (2019).
56. Spaldin, N. A., Cheong, S.-W. & Ramesh, R. Multiferroics: Past, present, and future. *Phys. Today* **63**, 38 (2010).
57. Hestenes, M. R. & Stiefel, E. Methods of conjugate gradients for solving linear systems. *J. Res. Natl. Bur. Stand.* **49**, 409 (1952).
58. Mentink, J. H., Tretyakov, M. V., Fasolino, A., Katsnelson, M. I. & Rasing, T. Stable and fast semi-implicit integration of the stochastic Landau–Lifshitz equation. *J. Phys. Condens. Matter* **22**, 176001 (2010).
59. Chen, Z. *et al.* 180° Ferroelectric Stripe Nanodomains in BiFeO3 Thin Films. *Nano Lett.* **15**, 6506–6513 (2015).
60. Chu, Y.-H. *et al.* Nanoscale Domain Control in Multiferroic BiFeO3 Thin Films. *Adv. Mater.* **18**, 2307–2311 (2006).
61. Huyan, H., Li, L., Addiego, C., Gao, W. & Pan, X. Structures and electronic properties of domain walls in BiFeO3 thin films. *Natl. Sci. Rev.* **6**, 669–683 (2019).
62. Lubk, A., Gemming, S. & Spaldin, N. A. First-principles study of ferroelectric domain walls in multiferroic bismuth ferrite. *Phys. Rev. B* **80**, 104110 (2009).
63. Diéguez, O., Aguado-Puente, P., Junquera, J. & Íñiguez, J. Domain walls in a perovskite oxide with two primary structural order parameters: First-principles study of $\mathrm{BiFeO}_{3}$. *Phys. Rev. B* **87**, 024102 (2013).
64. Ren, W. *et al.* Ferroelectric Domains in Multiferroic ${\mathrm{BiFeO}}_{3}$ Films under Epitaxial Strains. *Phys. Rev. Lett.* **110**, 187601 (2013).
65. Wang, Y. *et al.* ${\mathrm{BiFeO}}_{3}$ Domain Wall Energies and Structures: A Combined Experimental and Density Functional $\mathrm{\text{Theory}}\mathbf{+}\mathbf{U}$ Study. *Phys. Rev. Lett.* **110**, 267601 (2013).
66. Xue, F., Gu, Y., Liang, L., Wang, Y. & Chen, L.-Q. Orientations of low-energy domain walls in perovskites with oxygen octahedral tilts. *Phys. Rev. B* **90**, 220101 (2014).
67. Kirkpatrick, S., Gelatt, C. D. & Vecchi, M. P. Optimization by Simulated Annealing. *Science* **220**, 671–680 (1983).
68. Pinheiro, M., Ge, F., Ferré, N., Dral, P. O. & Barbatti, M. Choosing the right molecular machine learning potential. *Chem. Sci.* **12**, 14396–14413 (2021).
69. Geiger, M. & Smidt, T. e3nn: Euclidean Neural Networks. (2022) doi:10.48550/arXiv.2207.09453.
70. Paszke, A. *et al.* PyTorch: An Imperative Style, High-Performance Deep Learning Library. in *Advances in Neural Information Processing Systems* vol. 32 (Curran Associates, Inc., 2019).
71. Fey, M. & Lenssen, J. E. Fast Graph Representation Learning with PyTorch Geometric. *arXiv.org* https://arxiv.org/abs/1903.02428v3 (2019).
72. Momma, K. & Izumi, F. VESTA: a three-dimensional visualization system for electronic and structural analysis. *J. Appl. Crystallogr.* **41**, 653–658 (2008).
73. Hunter, J. D. Matplotlib: A 2D Graphics Environment. *Comput. Sci. Eng.* **9**, 90–95 (2007).
74. García-Palacios, J. L. & Lázaro, F. J. Langevin-dynamics study of the dynamical properties of small magnetic particles. *Phys. Rev. B* **58**, 14937–14958 (1998).
75. Kresse, G. & Hafner, J. Ab initio molecular dynamics for liquid metals. *Phys. Rev. B* **47**,

558 (1993).

76. Kresse, G. & Hafner, J. Ab initio molecular-dynamics simulation of the liquid-metal--amorphous-semiconductor transition in germanium. *Phys. Rev. B* **49**, 14251–14269 (1994).
77. Kresse, G. & Furthmüller, J. Efficiency of ab-initio total energy calculations for metals and semiconductors using a plane-wave basis set. *Comput. Mater. Sci.* **6**, 15–50 (1996).
78. Blöchl, P. E. Projector augmented-wave method. *Phys. Rev. B* **50**, 17953 (1994).
79. Perdew, J. P., Burke, K. & Ernzerhof, M. Generalized Gradient Approximation Made Simple. *Phys. Rev. Lett.* **77**, 3865–3868 (1996).
80. Ma, P.-W. & Dudarev, S. L. Constrained density functional for noncollinear magnetism. *Phys. Rev. B* **91**, 054420 (2015).
81. Dudarev, S. L., Botton, G. A., Savrasov, S. Y., Humphreys, C. J. & Sutton, A. P. Electron-energy-loss spectra and the structural stability of nickel oxide: An LSDA+U study. *Phys. Rev. B* **57**, 1505–1509 (1998).
82. Kornev, I. A., Lisenkov, S., Haumont, R., Dkhil, B. & Bellaiche, L. Finite-Temperature Properties of Multiferroic ${\mathrm{BiFeO}}_{3}$. *Phys. Rev. Lett.* **99**, 227602 (2007).
83. Kingma, D. P. & Ba, J. Adam: A Method for Stochastic Optimization. arXiv:1412.6980 (2017).
84. Elfwing, S., Uchibe, E. & Doya, K. Sigmoid-weighted linear units for neural network function approximation in reinforcement learning. *Neural Netw.* **107**, 3–11 (2018).

## Acknowledgments

This work is supported by NSFC (Grant Nos. 11825403, 12188101, and 11804138).

## Author contributions

H.Y.Y. conceived the model architecture, implemented the software, generated the datasets, ran experiments, and wrote the first version of the manuscript. Y.Z. contributed to the discussion of the code implementation. L.L.H. contributed to the discussion of spin-lattice dynamics and related LAMMPS code implementation. C.S.X. contributed to the discussion about simulations of $BiFeO_3$. R.W. contributed to the building of the domain wall structures of $BiFeO_3$. X.G.G. contributed to the discussion of the methods. H.J.X. supervised and guided the project from conception to design of experiments, methods, implementation, and dataset generation from effective Hamiltonian as well as analysis of data and results. All authors discussed the results and contributed to the revision of the manuscript.

## Competing interests

The authors declare no competing interests.

## Additional information

**Correspondence** and requests for materials should be addressed to Hongjun Xiang.